\documentclass[twocolumn]{jpsj3}

\title{DC Current Driven Critical Current Variation in Sr$_2$RuO$_4$-Ru Junction Proved
by Local Transport Measurements}

\author{Hiroshi Kambara\thanks{E-mail address: h.kambara@aist.go.jp}, Tetsuro Matsumoto,
Hiromi Kashiwaya, Satoshi Kashiwaya,
Hiroshi Yaguchi$^{1}$, Yasuhiro Asano$^{2}$, Yukio Tanaka$^{3}$, and Yoshiteru Maeno$^{4}$
}

\inst{National Institute of Advanced Industrial Science and Technology (AIST), Tsukuba 305-8568 \\
$^{1}$Department of Physics, Faculty of Science and Technology, Tokyo University of Science,
Noda 278-8510 \\
$^{2}$Department of Applied Physics, Hokkaido University, Sapporo 060-8628 \\
$^{3}$Department of Applied Physics, Nagoya University, Nagoya 464-8603 \\
$^{4}$Department of Physics, Kyoto University, Kyoto 606-8502
}

\abst{To search for new evidence of the chiral $p$-wave superconducting domain in Sr$_2$RuO$_4$,
we investigated the unconventional local transport characteristics of a microfabricated
Sr$_2$RuO$_4$-Ru eutectic junction.
We found that the anomalous hysteresis in voltage-current characteristics [as reported
in H. Kambara {\it et al.}: Phys. Rev. Lett. {\bf 101} (2008) 267003.] is strongly affected
by dc currents, but not by magnetic fields.
This suggests that dc current acts as a driving force to move chiral $p$-wave domain walls;
a domain wall trapped at a pinning potential is forced to shift to the next stable position,
thereby forming a larger critical current path and resulting in the anomalous hysteresis.}

\kword{Sr$_2$RuO$_4$, $p$-wave superconducting junction, eutectic, critical current, chiral domain, hysteresis}

\begin{document}
\maketitle

\section{Introduction}
\begin{figure}[ht]
\begin{center}
\includegraphics[width=1.0\linewidth]{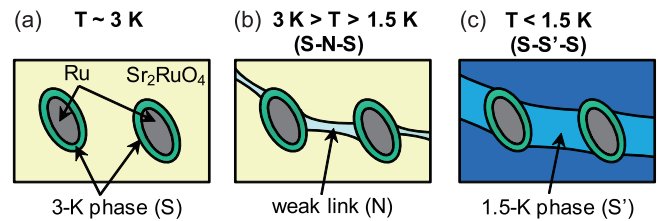}
\caption{\label{fig1}
(Color online)
Schematic image of superconducting channels in Sr$_2$RuO$_4$-Ru eutectic system:
(a) nucleation of 3-K phase superconductivity around Ru inclusions at 3 K,
(b) formation of weak links as S--N--S junctions at 3 K $> T >$ 1.5 K, and
(c) transformation of junctions to S--S$^{\prime}$--S junctions at $T <$ 1.5 K.}
\end{center}
\end{figure}

Layered perovskite Sr$_2$RuO$_4$ (SRO) \cite{maenoNT} is now widely believed to be one of
rare examples of odd-parity spin-triplet pairing superconductors \cite{maenoPT,mackenzie}.
Its superconducting order parameter is represented as the so-called chiral $p$-wave state
with broken time-reversal symmetry, which is similar to the superfluid A-phase of $^3$He \cite{tilley}.
The pure SRO phase has a sharp superconducting transition temperature ($T_c$) of 1.5 K (1.5-K phase),
while the SRO-Ru eutectic system shows a remarkable enhancement of $T_c$ onset up to 3 K with
its resistivity gradually decreasing to zero at 1.5 K. The superconductivity observed
at 3 K is called the 3-K phase \cite{maeno,ando,yaguchi}.
The 3-K phase is regarded as the interface superconductivity in the SRO region between SRO
and Ru inclusions because its superconducting volume fraction has been found to be considerably
smaller than that of the pure 1.5-K phase from specific heat measurement \cite{yaguchi}.
Tunnel junction experiments using Ru interfaces \cite{maoPRL,kawamura} showed a zero-bias conductance peak
due to Andreev resonance, \cite{tanaka} suggesting that the 3-K phase shows non-$s$-wave superconductivity.
Recently, we have performed local transport measurements on microfabricated samples of
the SRO-Ru eutectic system and observed the development of the superconducting linkage channels
connecting Ru inclusions with decreasing temperature below 3 K \cite{kambara}.
From the measurement of the critical current ($I_c$) in the linkage channels, we found that
the 3-K phase also has odd-parity pairing symmetry, similar to the 1.5-K phase \cite{kambara}.
This shows that the SRO-Ru eutectic system includes naturally formed $p$-wave superconducting junctions
in itself depending on temperature;
the 3-K phase region--normal-state region--3-K phase region as
the superconductor(S)--normal-metal(N)--S junction at 3 K $> T >$ 1.5 K,
while the 3-K phase region--1.5-K phase region--3-K phase region as
the S--S$^{\prime}$--S junction at $T <$ 1.5 K (Fig. \ref{fig1}).

We observed quite anomalous hysteresis in the voltage--current ($V-I$) and
differential-resistance--current ($dV/dI-I$) characteristics of these junctions
below 2 K \cite{kambara}.
``Anomalous'' hysteresis implies that it is different from the usual hysteresis observed in usual Josephson
junctions (JJs). That is, (i) voltage discontinuously {\it decreases} when the $V-I$ curve
switches to the next branch with increasing current, and therefore, (ii) the hysteresis loop shows
the {\it opposite} direction compared to usual JJs \cite{barone}.
Thus, these features indicate the emergence of the internal degrees of freedom of the superconducting state.
One possible explanation of the anomalous hysteresis is due to the existence of chiral $p$-wave domains
reflecting the chiral $p$-wave pairing of SRO.
The existence of chiral domains in SRO has been suggested by several experiments;
the domain size has been estimated to be approximately 50--100 $\mu$m from Kerr effect
measurement \cite{xia} and approximately $1$ $\mu$m from a SRO-Cu-Pb junction experiment \cite{kidwingira}.

To investigate such anomalous transport properties in detail and examine chiral domains,
we measured the local transport characteristics of microfabricated SRO-Ru eutectic samples,
both in-plane and out-of-plane directions.
The advantage of milling a sample down to the micrometer scale by a focused ion beam (FIB) is
that one can extract each conducting channel from a complex network \cite{hooper} and detect
anomalous features without averaging over the bulk sample. 
In this paper, we report that the anomalous hysteresis is strongly affected by dc currents,
but not affected by magnetic fields.
This suggests a dc-current-driven $I_c$ variation, which is explained by the discontinuous motion of
the chiral domain wall trapped at a pinning potential, which is discussed below.

\section{Experiment}
Eutectic crystals of SRO-Ru were grown in an infrared image furnace
by the floating zone method \cite{mao}.
The sample preparation is as follows.
First, a thin platelet of a crystal was prepared by cleaving a crystal along the $ab$-plane and mechanically
polishing it down to a thickness of a few tens of micrometers along the $c$-axis.
Second, the platelet was roughly cut into a small piece of submillimeter width in the $ab$-plane.
Third, the small piece was glued on a SrTiO$_3$ substrate using epoxy adhesive.
Last, four contacts for transport measurement were formed on the piece using an evaporated gold film.
Then, the sample was milled by an FIB to form a ``microbridge'' between the voltage lead contacts
(Fig. \ref{fig2}(a)).
Since electric fields concentrate in the microbridge (typical size is shown in Fig. \ref{fig2}(b)),
the measured electric resistance is predominated by the electric resistance of the microbridge.
Thus, we can study local transport properties by focusing on the microbridge from a complex network
of linkage channels spread over the entire sample.
To measure transport characteristics along $I//c$, we formed a crank-shaped path by milling
two alternate slits along the $c$-axis by an FIB (Figs. \ref{fig2}(c) and \ref{fig2}(e)) in a manner
similar to the fabrication of intrinsic JJs of high-$T_c$ cuprates \cite{matsumotoFIB,hkashiwaya}.
We confirmed the existence of Ru inclusions in the microbridge region by FIB milling
after the transport measurements were done (Fig. \ref{fig2}(f)).
Figure \ref{fig2}(f) shows that a piece of Ru inclusion was embedded slightly below the top surface at the
center of the microbridge. We observed about six pieces of Ru inclusions including small ones
between the $c$-axis slits.
The transport measurements were performed by dc and ac methods. In the ac method, ac modulation
with an amplitude of 30 $\mu$A$_{\rm {rms}}$ and a frequency of 977 Hz was used.
The sample was cooled below 1 K using a $^3$He cryostat. External magnetic fields were applied
parallel to the $c$-axis for evaluating the magnetic field effects.

\section{Results}
\subsection{$R-T$ curves of FIB-milled samples}
\begin{figure}[ht]
\begin{center}
\includegraphics[width=1.0\linewidth]{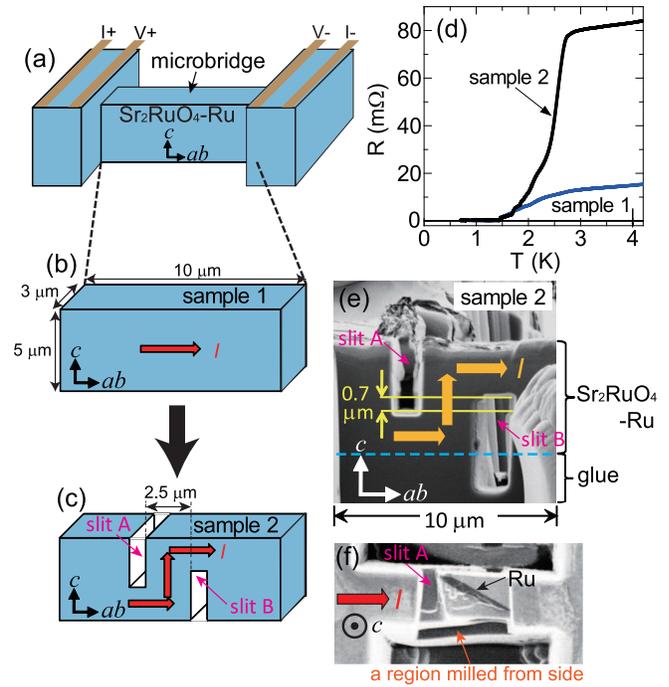}
\caption{\label{fig2}
(Color online)
(a) Schematic sample configuration for local transport measurements (not drawn to scale), and
configurations of microbridges of (b) sample 1 for $I//ab$ and (c) sample 2 for $I//c$, respectively.
Sample 2 was made from sample 1 by milling slits along the $c$-axis.
(d) Zero-bias $R$ vs. $T$ for samples 1 and 2.
(e) Scanning ion microscope (SIM) image ($10 \times 10$ $\mu$m$^2$) from the side view of sample 2.
(f) SIM image from the top view obtained after the top and side surfaces were slightly milled.
A piece of Ru inclusion ($\sim3$ $\mu$m length) was embedded at the center of the microbridge.}
\end{center}
\end{figure}

Figures \ref{fig2}(b) and \ref{fig2}(c) show schematic sample configurations
for $I//ab$ (sample 1) and $I//c$ (sample 2), respectively
(note: sample 1 is the same as sample D used in ref. 11). 
In sample 2, we note that the resistance of the $I//c$-part is high and that of the $I//ab$-part is low
($\sim$18\%) at 4.2 K. In this study, we regard the configuration of sample 2 as $I//c$.
Figure \ref{fig2}(d) shows the differential resistance at zero-bias current ($R$) versus temperature ($T$)
curves of samples 1 and 2.
If we assume that the overlap path length (0.7 $\mu$m) along the $c$-axis only contributes to the $c$-axis
resistivity, the ratio of out-of-plane resistivity ($\rho_c$) to in-plane resistivity
($\rho_{ab}$) of this material is estimated to be $\rho_{c}/\rho_{ab} \simeq 40$ at 4.2 K.
This is comparable to previously reported data \cite{ando}.

\subsection{Anomalous $V-I$ and $dV/dI-I$ characteristics}
\begin{figure}[ht]
\begin{center}
\includegraphics[width=1.0\linewidth]{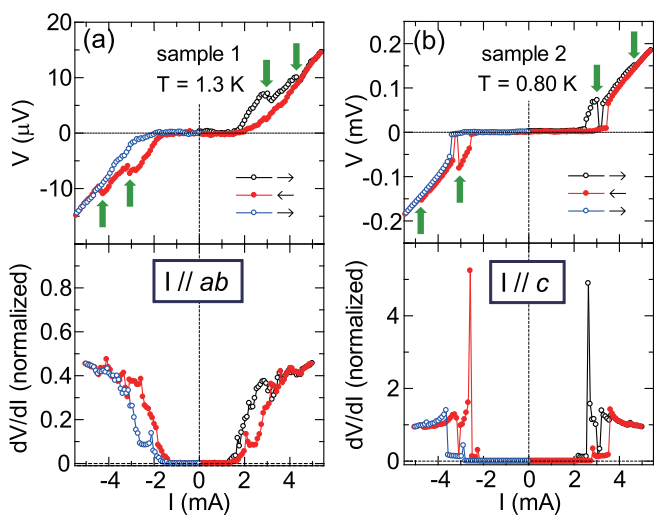}
\caption{\label{fig3}
(Color online)
Typical $V-I$ and $dV/dI-I$ characteristics with anomalous hysteresis of (a) sample 1
at $T$ = 1.3 K and (b) sample 2 at $T$ = 0.80 K.
The $dV/dI$ values were normalized to the zero-bias $dV/dI$ at 4.2 K.
The bold arrows in $V-I$ characteristics denote the anomalous voltage changes.
The open and filled symbols denote the different sweep directions, from zero to maximum (open black),
maximum to minimum (solid red), and minimum to zero (open blue).}
\end{center}
\end{figure}

Figures \ref{fig3}(a) and \ref{fig3}(b) show typical $V-I$ and $dV/dI-I$ characteristics of samples 1
(Figs. 4(a) and 4(b) in ref. 11) and 2 obtained by the dc and ac methods, respectively.
The most peculiar feature of these systems is that the absolute values of voltage abruptly decrease
at the thresholds
$I_{\rm {th1}} = \pm3$ mA and $I_{\rm {th2}} = \pm4.3$ mA for sample 1, and at
$I_{\rm {th1}} = \pm3.1$ mA and $I_{\rm {th2}} = \pm4.7$ mA for sample 2
with increasing absolute values of dc currents, which are shown by bold arrows in the $V-I$ characteristics
of Figs. \ref{fig3}(a) and \ref{fig3}(b).
For the ac method, since the $dV/dI-I$ peaks or steps correspond to the $I_c$s of the superconducting
linkage channels \cite{kambara},
we can see that $I_c$ becomes larger with decreasing current rather than increasing current.
Furthermore, no negative $dV/dI-I$ value means that the unusual voltage decrease is a switching phenomenon
in which no intermediate states exist.
Thus, we found that the anomalous hysteresis is a common feature observed not only in the
in-plane direction but also in the out-of-plane direction.
The peaks in $dV/dI-I$ of $I//c$ (sample 2) are considerably sharper than those of $I//ab$ (sample 1)
reflecting a quasi-two-dimensional conductor; a junction behaves like a tunnel along the $c$-axis;
on the other hand, it behaves as a weak link in the $ab$-plane.
We note that the reproducibility of the hysteresis after thermal cycling (warming up to 300 K
and cooling again) was amazingly high up to around 10 cycles for both samples 1 and 2.
Furthermore, we note that the anomalous hystereses were observed for most of the samples of SRO-Ru.
Thus, the anomalous hysteretic feature is not an incidental phenomenon, but an intrinsic property.

\subsection{Dc current sweep width dependence of hysteresis}
\begin{figure}[ht]
\begin{center}
\includegraphics[width=1.0\linewidth]{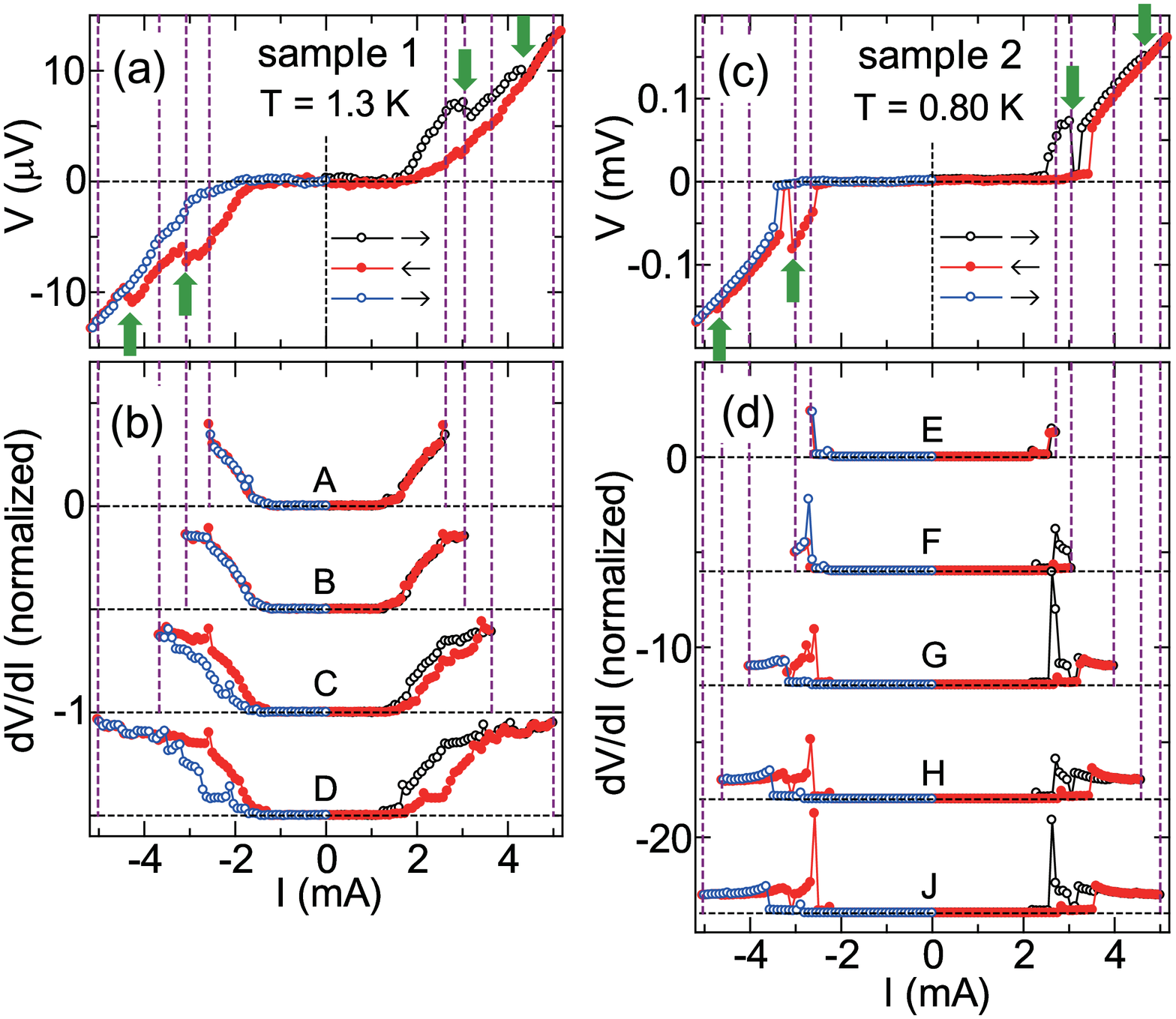}
\caption{\label{fig4}
(Color online)
Dc current sweep width dependence of hysteresis for sample 1 ((a) and (b)) and sample 2 ((c) and (d)).
(a), (c) $V-I$ characteristics as a reference.
(b) $dV/dI-I$ characteristics of sample 1 for (A) $\pm$2.6, (B) $\pm$3.1, (C) $\pm$3.7, and
(D) $\pm$5.0 mA current sweeps.
(d) $dV/dI-I$ characteristics of sample 2 for (E) $\pm$2.7, (F) $\pm$3.0, (G) $\pm$4.0,
(H) $\pm$4.6, and (J) $\pm$5.0 mA current sweeps.
The vertical dashed lines represent the sweep ranges.
The open and filled symbols denote the different sweep directions, which are the same as those
in Fig. \ref{fig3}.
The curves are offset by -0.5 units in (b) and by -6 units in (d) for clarity.}
\end{center}
\end{figure}

Next, we show that the shape of the hysteresis strongly depends on the dc current sweep width.
Figures \ref{fig4}(b) and \ref{fig4}(d) show $dV/dI-I$ characteristics for different sweep widths
of samples 1 (A--D) and 2 (E--J), respectively.
For sample 1, no hysteresis was observed in (A and B). On the other hand in (C), a narrow hysteresis,
compared with that in (D), was observed.
For sample 2, no hysteresis was observed in (E) and a negative-current region in (F).
On the other hand in (G), a narrow hysteresis, compared with those in (H and J), was observed.
Note that a small difference between the hystereses of (G) and (H and J) appears when current is decreased.
Also note that the shapes of the hystereses of (H and J) are almost the same except for the $dV/dI-I$ peak
heights. The experimental results show that hysteresis width increases in a steplike manner
when currents larger than $I_{\rm {th}}$s (shown by bold arrows in Figs. \ref{fig4}(a) and
\ref{fig4}(c)) are applied. We note that the difference in the presence of hysteresis in (F),
i.e., hysteresis in the positive-current region, whereas no hysteresis in the negative-current region,
is due to a small fluctuation in $I_{\rm {th}}$.
For the same reason, the curve (H) shows a large hysteresis, which is similar to (J) rather than to (G),
though the current sweep width of (H) is slightly small compared with $I_{\rm {th2}}$ in the $V-I$ curve.

\subsection{Magnetic field effect on hysteresis}
\begin{figure}[ht]
\begin{center}
\includegraphics[width=1.0\linewidth]{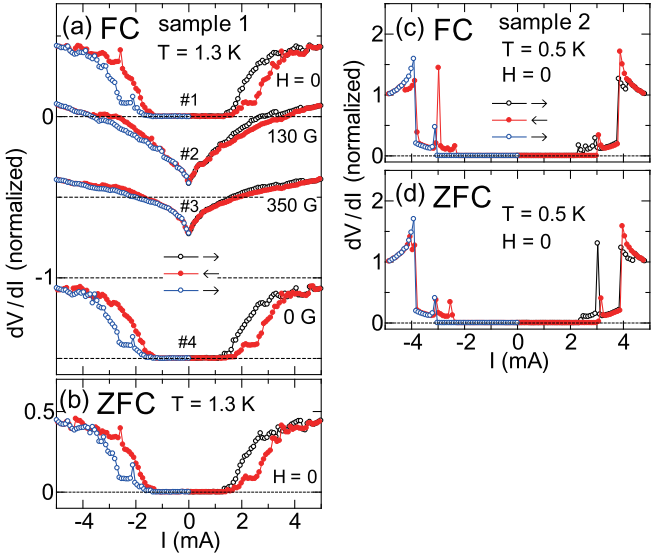}
\caption{\label{fig5}
(Color online)
Normalized $dV/dI-I$ curves obtained after (a) FC in 130 G parallel to $c$-axis and
(b) ZFC for sample 1 measured at 1.3 K, and those obtained after (c) FC in 1000 G parallel to $c$-axis and
(d) ZFC for sample 2 measured at 0.5 K.
In (a), the $dV/dI-I$ curves obtained at $H$ = 130, 350, and 0 G after FC are shown together.
The curves are offset by -0.5 units for clarity.
The numbers denote the measured sequence.
The open and filled symbols denote the different sweep directions, which are the same as those
in Fig. \ref{fig3}.}
\end{center}
\end{figure}

To investigate the magnetic field effect, magnetic field cooling (FC) and zero-field cooling (ZFC)
were carried out. If the anomalous hysteresis originates from a magnetic vortex, the shape of the
hysteresis curve would change between FC and ZFC.
For FC, first, we applied a magnetic field larger than the lower critical field
$H_{c1}$ ($\approx$ 70 G \cite{deguchi}) parallel to the $c$-axis at 4.2 K
(130 G for sample 1 and 1000 G for sample 2), and cooled the samples below 1 K
under the magnetic fields.
Next, we switched off the fields and measured $dV/dI-I$ at zero field.
On the other hand, for ZFC, we cooled the samples under zero field and measured $dV/dI-I$
at zero field.
Figures \ref{fig5}(a) and \ref{fig5}(b) (\ref{fig5}(c) and \ref{fig5}(d)) show the $dV/dI-I$ curves
of FC and ZFC for sample 1 (2), respectively.
There was no significant difference between the $dV/dI-I$ curves of FC and ZFC, for both the samples.
Figure \ref{fig5}(a) also shows the data obtained under magnetic fields applied after FC.
In any case, when the applied magnetic fields are higher than $H_{c1}$, magnetic vortices can penetrate
the samples. If trapped vortices make $I_c$ variation, a difference in the shape of the hysteresis
between the $dV/dI-I$ curves of FC and ZFC should be expected.
However, no difference is observed between the $dV/dI-I$ curves obtained at zero field
before and after the application of the magnetic fields.
These results imply that the anomalous hysteresis is not caused by the magnetic vortex effect.
In contrast to the dc current effect, magnetic field showed no significant effect on the
anomalous hysteresis.

\section{Discussion}
The experimental results are summarized as follows.
(i) Anomalous hystereses appear for both current directions ($I//ab$ and $I//c$).
(ii) The width of the hysteresis curves changes discontinuously depending on current sweep width. 
(iii) Magnetic fields do not affect the hysteresis.

Next, we analyze these experimental results by using the chiral domain model proposed
in our previous study \cite{kambara}.
Using this model, we assume that the angular momentum transfer from a given chiral state
($p_x - ip_y$) to the opposite chiral state ($p_x + ip_y$) by dc current induces domain wall (DW) motion.
Thus, dc current causes an unusual $I_c$ variation as a function of position because
$I_c$ is expected to be proportional to the cross-sectional area of spatially inhomogeneous linkage channels.
If a DW exists, it is preferentially trapped at a defect at the beginning
because a small overlapping area is energetically favorable for different order parameters
of antiparallel domains. Thus, $I_c$ would increase with increasing dc currents and cause
the anomalous hysteresis.
We also note that there is no DW between parallel domains. Thus, the anomalous hysteresis
would appear only between antiparallel domains.

Regarding feature (ii), if we assume that $I_{{\rm th}}$ originates from a pinning potential due to,
for example, a type of lattice defect, the well-reproducible $I_{{\rm th}}$ after thermal cycle is explainable.
Assuming that applying dc currents reduces effective pinning potential, the trapped DW
can escape when sufficiently large dc currents, which can overcome the pinning potential, are applied.
This is analogous to the depinning phenomenon of charge density waves with electric fields exceeding
a certain threshold value \cite{gruner}.
Using the simplest model in this case, we consider a tilted washboard potential as the pinning potential,
although the pinning sites are not aligned periodically in an actual case.
Then, the effective potential $U_{\rm {eff}}$ is given as a function of the position $x$ as
$U_{\rm {eff}} \approx -U \cos k x -\alpha I x$,
where $U$ is the amplitude of the potential corrugation and $k$ and $\alpha$ are numerical constants.

\begin{figure}[ht]
\begin{center}
\includegraphics[width=0.85\linewidth]{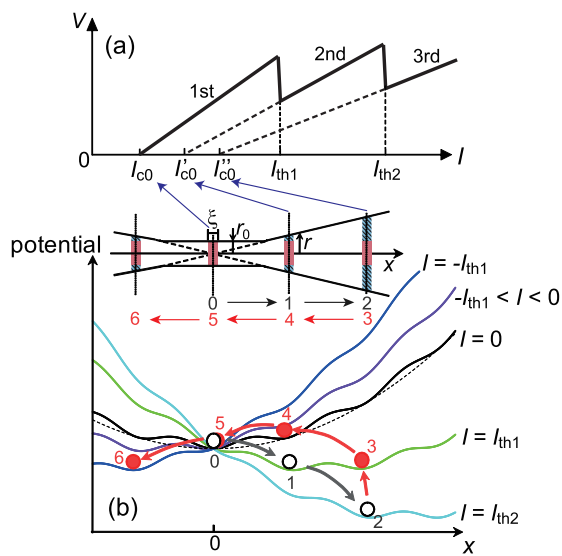}
\caption{\label{fig6}
(Color online)
(a) Schematic $V-I$ characteristics and (b) model of effective pinning potential of
domain wall as a function of position of channel under dc currents flowing conditions.
The open and solid (red) circles denote the trapped positions at the local minima with
increasing and decreasing currents, respectively.
The dashed quadratic curve shows the condensation energy loss.
Inset: model of chiral domain trapped in horn-shaped channel.
$I_{c0}$ increases ($I_{c0}'$, $I_{c0}''$) with the cross-sectional area of the channel.
The hatched light blue parts of the cross section of the domain wall correspond to the volumes of
the condensation energy loss.}
\end{center}
\end{figure}

We should consider the variation in the superconducting condensation energy due to the variation
in the cross-sectional area of the channel at the pinning sites.
For antiparallel domains, since a small cross-sectional area is energetically favored, an increase in
the cross-sectional area means a {\it loss} of the condensation energy.
Therefore, the energy increase is proportional to the volume $\pi (r^2 - r_0^2) \xi$,
assuming a simple ``horn''-shaped channel (inset of Fig. \ref{fig6}(b)), where
$r_0$ is the radius of the channel at the origin ($x = 0$), which has a minimum cross section,
$r$ is the radius of the channel away from the origin, and
$\xi$ is the coherence length of SRO.
Assuming that $r$ is proportional to $x$ away from $x = 0$ in the simplest case, i.e., $r = \beta x$
($\beta$ is a numerical constant),
$U_{\rm {eff}}$ is rewritten as a function of $x$ (without a constant term) as
$U_{\rm {eff}} \approx -U \cos k x -\alpha I x + (\mu_{0} H_{c}(0)^{2}/2) \pi \beta^2 x^2 \xi$,
where $\mu_{0} H_{c}(0)^{2}/2$ is the condensation energy per unit volume.
We have roughly estimated the order of magnitude of the 3rd term, i.e., the condensation energy loss,
to be $\sim$0.1--1 eV, where $\mu_0 H_c(0)$ = 0.023 T \cite{mackenzie}; $\xi$ = 66 nm (in-plane),
3.3 nm (out-of-plane) at $T$ = 0 K for the 1.5-K phase \cite{mackenzie};
$x \approx$ 1 $\mu$m; and $\beta \approx$ 1/20 from a rough estimation of the ratio of the cross sections.
On the other hand, we estimated $U$ to be $\sim$0.1 eV for an order of magnitude assuming that
$U$ is comparable to the typical pinning potential of a magnetic vortex \cite{tinkham}.
Thus, these two terms are comparable.
  
Figure \ref{fig6}(b) shows examples of $U_{\rm {eff}}$ variation under dc currents for certain
values of parameters.
If we assume viscous movement for the DW, it will be trapped at the local potential minima.
Therefore, the DW moves to the next local potential minimum at $I_{\rm {th1}}$ and $I_{\rm {th2}}$
with increasing dc currents from zero (process: 0$\rightarrow$1$\rightarrow$2).
On the other hand, reducing dc currents from $I_{\rm {th2}}$ to $I_{\rm {th1}}$ (process: 2$\rightarrow$3)
does not change the position of the DW.
As a result, a larger $I_c$ is observed during decreasing dc current process.
An important feature is that the DW moves back to the origin around $I \approx 0$
due to the quadratic potential.
Thus, the hysteresis loop starts again from a lower $I_c$ even in the region of negative currents,
which is similar to the case of positive currents, and repeats the cycle.

Generally in a chiral domain, edge currents are expected to flow along DWs \cite{matsumoto,kwon}.
Thus, it is naturally expected that magnetic fields would interact with the magnetic moment
of the edge currents.
Feature (iii), however, seems to contradict this speculation.
One possible reason for this inconsistency is that the actual chiral domain is not a simple
$p_x \pm ip_y$ state, but the coexistence of the $p_x \pm ip_y$ and $p_x$($p_y$) states.
Because of the translational symmetry breaking at sample edges, the $p_x$ state, which is expected for
the 3-K phase symmetry \cite{sigrist}, seems to be stabilized near the sample edges.
In fact, we observed a rough tendency for the ratio of the 3-K phase to the 1.5-K phase to increase
with successive FIB processing (smaller than $\sim$10 $\mu$m in sample size) in $R-T$ measurements
for various samples.
This suggests that the coexistence state should be taken into account as the microbridge becomes smaller.
If the $p_x$ state forms along the edge, the effect of the chiral edge currents would be reduced
by Meissner shielding current flowing along the outside of the $p_x \pm ip_y$ state.
As a result, the magnetic field effect on the anomalous hysteresis would weaken or disappear
because of the mixing of the time-reversal symmetry conserving states.
This speculation is consistent with experimental results in which no spontaneous supercurrents
are observed using a scanning Hall probe or a superconducting quantum interference device
(SQUID) \cite{bjornsson,kirtley}.

From feature (i) it is found that, in the chiral domain scenario, the DW can move not only in the
in-plane direction but also in the out-of-plane direction, although the chiral domain is usually
discussed only in the context of the in-plane direction theoretically.
Further experimental work is necessary to validate the existence of chiral domains
or to determine whether other models can be applied to the SRO-Ru system.

\section{Conclusion}
In summary, we have observed anomalous transport characteristics in microfabricated samples
of the Sr$_2$RuO$_4$-Ru eutectic system, both in-plane and out-of-plane directions.
The anomalous hysteresis in differential-resistance and voltage--current characteristics
are explained by the presence of chiral domains. The domain wall, which is trapped at
lattice defects, is driven by large currents beyond the pinning potential.
The absence of the magnetic field effect on the hysteresis implies the coexistence of the $p_x \pm ip_y$
and $p_x$($p_y$) states in a microbridge sample with a size of the order of a few micrometers.
All the observed features suggest the emergence of the internal degrees of freedom of
the chiral $p$-wave state.

%

\section*{Acknowledgments}
We are very grateful to K. Shirai for providing technical support and to H. Shibata and
S. Kawabata for helpful discussions.
This work was financially supported by a Grant-in-Aid for Scientific Research on Priority
Areas (No.~17071007) from MEXT and Grants-in-Aid for Scientific Research (c) (No.~20540392)
and for Young Scientists (B) (No.~21740276) from JSPS, Japan.

\end{document}